\documentclass{PoS}

\title{Excited mesons from $N_f=2$ dynamical Clover Wilson lattices}

\ShortTitle{Excited mesons from $N_f=2$ dynamical Clover Wilson lattices}

\author{Tommy~Burch, \speaker{Christian~Hagen}, and Andreas~Sch\"afer\\

	Institut f\"ur Theoretische Physik\\
        Universit\"at Regensburg\\
	D-93040 Regensburg, Germany.\\

        E-mail: \email{christian.hagen@physik.uni-regensburg.de}\\
       }

\abstract{
We study mesons on the lattice with a special focus on excited states.
For that purpose we construct several quark sources with different
spatial smearings, including p-waves. These quark sources are then
combined with the appropiate Dirac structures to form meson interpolators
of definite spin. We use these operators to construct a cross correlation
matrix from which we extract ground and excited meson states using the
variational method. For the calculations we use gauge configurations with
$N_f=2$ dynamical Clover Wilson fermions provided by the CP-PACS
collaboration. We show preliminary results for pseudoscalar, scalar, vector 
and pseudovector mesons.
}

\FullConference{XXIVth International Symposium on Lattice Field Theory\\
                July 23-28, 2006\\
                Tucson, Arizona, USA}

\begin{document}

\section{Introduction}

Ground state spectroscopy in lattice calculations appears to be well understood. Excited state
spectroscopy, however, is still a challenging task. There are two major problems: First, one has to 
improve the overlap of the interpolating fields with excited states. Excited hadron states include both 
radial and orbital excitations and thus allowing for such excitations should be implemented in a lattice
calculation to obtain more realistic results. The second problem is finding reliable means to disentangle 
ground and excited states in the hadron spectrum. Our method of choice is a variational approach
\cite{Michael:1985ne,Luscher:1990ck}. It not only allows us to extract excited states but also separates 
physical states from ghost contributions in quenched and partially quenched calculations.

\section{Simulation details}

We perform lattice calculations with dynamical gauge configurations with $N_f=2$ Clover Wilson Fermions.
The configurations were generated by the CP-PACS collaboration, which made them publicly available. At the 
moment there are three lattices available: a $12^3\times24$, a $16^3\times32$, and a $24^3\times48$. All 
these lattices have approximately the same spatial volume of about $2.5$ fm, but differ in their lattice 
spacing. This enables us to perform a continuum extrapolation. For the preliminary results we are 
presenting here we have used 100 configurations per sea quark mass of the $16^3\times32$ lattice which has 
a lattice spacing of $0.1555(17)$ fm. For further information about the lattices and how they were 
generated, see Ref. \cite{Aoki:1999ff,AliKhan:2001tx}.

For our simulations, we make use of the Chroma software package \cite{Edwards:2004sx} from the USQCD. 
This package has the advantage that most of the applications needed in lattice QCD are already 
implemented and it can be easily installed on various platforms, including QCDOC.

\section{The Method}

To obtain reliable results for excited states, we follow a procedure which has already proven to be
very successful in quenched calculations \cite{Burch:2006dg,Burch:2006cc}. First, we generate several 
spatially different quark sources. In previous studies, we considered a narrow and a wide Gaussian 
source (in the following called $n,w$, respectively), obtained from gauge covariant smearing, to allow 
for a node in the radial wave function of the quark. Here we also include p-wave sources ($p_x$, $p_y$, 
$p_z$) which we generate by acting with a covariant derivative on the wide smeared source. This
enables us to not only explore the possibility of radial excitations, but also orbital ones. In addition 
to these sources we add a local source ($L$) so that we can examine physical matrix elements. 

We then combine these sources with the appropriate Dirac structures to obtain interpolating fields of 
definite spin. By doing so, we end up with a large number of interpolators for each meson channel 
(e.g., 15 different interpolators in the case of the pseudoscalar meson). From these interpolators we 
construct a matrix of correlators and then apply the variational method which has been proposed by 
Michael \cite{Michael:1985ne} and later refined by L\"uscher and Wolff \cite{Luscher:1990ck}. Here, one has 
to solve a generalized eigenvalue problem. This method has some advantages. First, the system has 
full freedom to choose relative contributions of the different interpolators in the diagonalization step.
Second, this approach can separate ghost contributions in quenched and partially quenched results, as has
been shown in Ref. \cite{Burch:2005wd}. In the case where one uses more than one local operator one can 
also extract ratios of couplings to excited states, see Ref. \cite{ehmann06}.

The masses of the excited states are obtained by looking at the eigenvalues, which to leading order 
behave as
\begin{eqnarray}
\label{singleexp}
\lambda^{(k)}(t) & \propto & e^{-t \, M_k}\; .
\end{eqnarray}
From these eigenvalues we then construct effective mass plots by using
\begin{eqnarray}
am_k^{eff}\left(t+\frac{1}{2}\right) & = & 
\ln\left(\frac{\lambda^{(k)}(t)}{\lambda^{(k)}(t+1)}\right) \; .
\end{eqnarray}
When we find a plateau in these plots and if the corresponding eigenvectors are steady we conclude 
that the signal of the considered state is disentangled from higher excitations. We then fit the
eigenvalue to a single exponential according to Eq.~(\ref{singleexp}). In the following we present 
the results of these fits for different mesons.

\section{Discussion of the Results}

\begin{figure}
\begin{center}
\resizebox{0.7\textwidth}{!}{\includegraphics[clip]{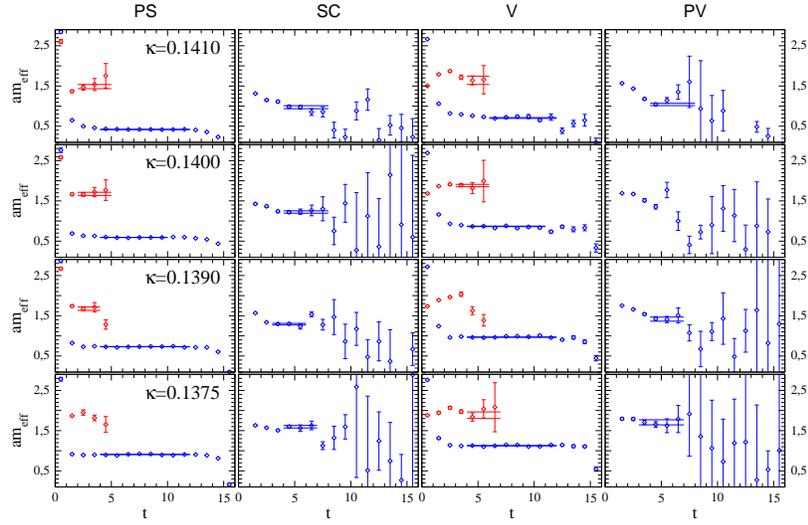}}
\end{center}
\caption{
Effective masses for pseudoscalar(PS), scalar(SC), vector(V), and pseudovector(PV), shown for 
completely degenerate quark masses. The horizontal lines denote the time intervals of our fits and
represent the fit results $m \pm \sigma_m$, where $\sigma_m$ is the statistical error obtained 
from single elimination jackknife.
}
\label{effmass}
\end{figure}

In Fig. \ref{effmass}, we show a collection of effective mass plots for the different meson channels
pseudoscalar(PS), scalar(SC), vector(V), and pseudovector(PV). The effective masses are shown only 
for the completely degenerate case
\begin{equation}
\kappa_{val}^{(1)} = \kappa_{val}^{(2)} = \kappa_{sea} = \kappa.
\end{equation}
With our limited statistics, we find good plateaus for most of the ground states. But the results 
for the excited states do not look encouraging. In fact, we are able to obtain a first excited 
state only in the PS and V channels. 

\begin{figure}
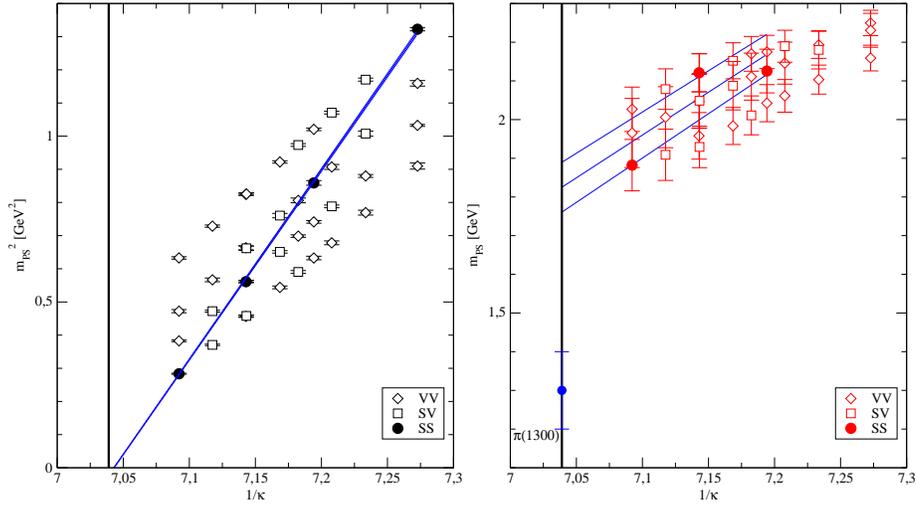

\begin{center}
\resizebox{0.8\textwidth}{!}{\includegraphics[clip]{pion_ground_16_final.eps} \includegraphics[clip]{pion_exc_16_final.eps}}
\caption{
Results for the pseudoscalar meson ground (left plot) and excited state (right plot) versus 
$1/\kappa$. Filled circles denote the SS case, open squares the SV case, and open diamonds the 
VV case. The vertical line is $1/\kappa_{crit}$ obtained by the CP-PACS collaboration. We also 
show a naive chiral extrapolation of the SS results. Also the experimental value is included.
}
\label{psresults}
\end{center}
\end{figure}

\begin{figure}
\begin{center}
\resizebox{0.7\textwidth}{!}{\includegraphics[clip]{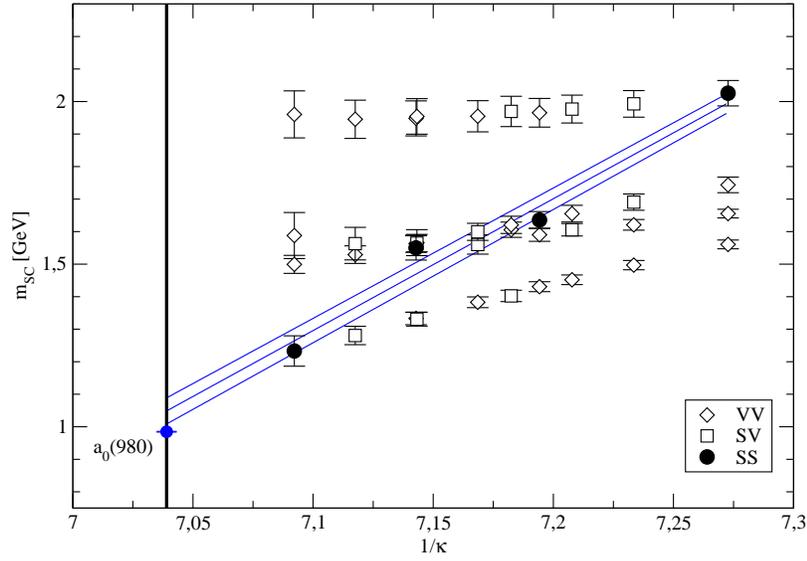}}
\caption{
Same as in Fig. 2, but for the scalar meson ground state.
}
\label{scresults}
\end{center}
\end{figure}

\begin{figure}
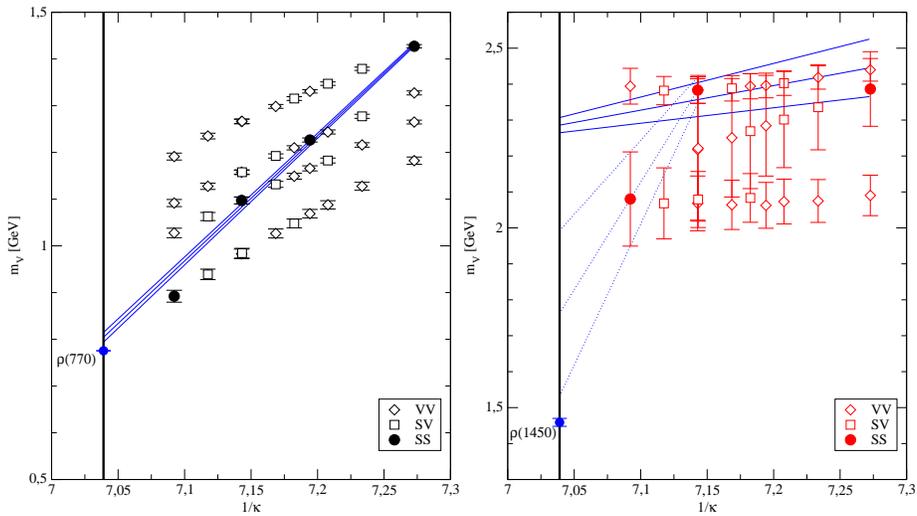

\begin{center}
\resizebox{0.8\textwidth}{!}{\includegraphics[clip]{vector_ground_16_final.eps} \includegraphics[clip]{vector_exc_16_final.eps}}
\caption{
Same as in Fig. 2, but for the vector meson ground and excited state. For the excited state we
tried two different chiral extrapolations.
}
\label{vresults}
\end{center}
\end{figure}

\begin{figure}
\begin{center}
\resizebox{0.7\textwidth}{!}{\includegraphics[clip]{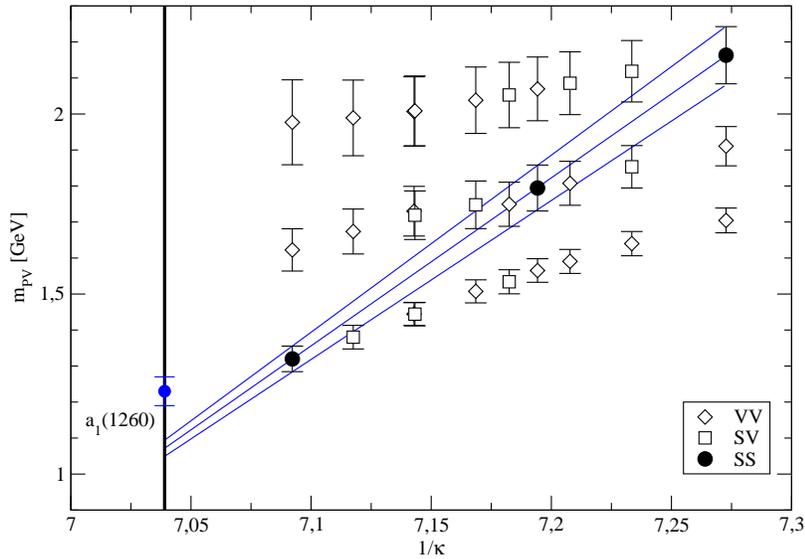}}
\caption{
Same as in Fig. 2, but for the pseudovector meson ground state.
}
\label{pvresults}
\end{center}
\end{figure}

In Fig. \ref{psresults}-\ref{pvresults}, we plot the results of our fits versus 
\begin{equation}
\frac{1}{\kappa} = \frac{1}{2} \left( \frac{1}{\kappa_{val}^{(1)}} + \frac{1}{\kappa_{val}^{(2)}} \right).
\label{invkappa}
\end{equation}
We destinguish three different cases: The case where both valence quark masses are equal to the 
sea quark mass, denoted as SS, the case where only one valence quark has the same mass as the sea 
quark, denoted as SV, and the VV case where both valence quarks have a mass which differs from the 
sea quark mass. In all the figures the vertical line represents $1/\kappa_{crit}$ obtained by the 
CP-PACS collaboration using all available configurations. Since our results are still very 
preliminary we perform only linear fits for the chiral extrapolation. 

In the left hand plot of Fig. \ref{psresults}, one can see that we obtain with our limited 
statistics a result for $\kappa_{crit}$ which is comparable to the CP-PACS result for this quantity.

In all the cases we cannot exclude systematic effects which are due to the limited statistics we 
have so far. It might, for example, be possible, that with larger statistics some of the plateaus in
Fig. \ref{effmass} change and a fit interval starting one timeslice later might be more appropriate.
Especially our excited states with their short plateaus can be strongly affected by such systematic 
shifts. This might also explain why our results for the excited pseudoscalar and vector meson are 
extrapolating to values which are much higher than the experimental ones.

Apart from this, we also find that the widths of our Gaussian sources depend on the sea 
quark mass. This makes it rather difficult to tune the smearing parameters to obtain widths 
in physical units which are approximately the same. What we also find is that a cross 
correlation matrix of interpolators built from $L$ and $n$ sources gives much better results 
than using interpolators built from $n$ and $w$ sources. This suggests, that our narrow 
and wide smeared sources are too similar to allow to disentagle ground
and excited states.

\section{Summary and outlook}

We have presented preliminary results of our calculation of ground and excited meson states from
dynamical configurations with $N_f=2$ Clover Wilson fermions. For all considered channels we find 
results for the ground states and in the pseudoscalar and vector channels we are even able to extract 
a first excited state. 

We encounter some difficulties to optimize the smearing parameters for our calculation since we 
find that the width of our Gaussian source depends on the sea quark mass. In addition to this,
we observe that interpolators built from these Gaussian sources are giving worse signals for
the excited states than interpolators built from one Gaussian and a local source. Possibly, this
means that the Gaussians used are too much alike. 

To circumvent this problem we are currently running simulations with a different set of sources.
We use a narrow source with much smaller width and replace our old wide source by the narrow source
upon which we apply a Laplacian (The Laplacian being a scalar operator does not change the
quantum numbers of the hadron).

Our future plans also include  performing the same calculations on the other lattices of the CP-PACS 
collaboration to be able to do a continuum extrapolation of our results. Especially the couplings
to excited states are in that respect of great interest.

\section{Acknowledgements}

We thank C. Gattringer, L. Ya. Glozman and C. B. Lang for interesting discussions and S. Solbrig for very
valuable hardware support. This work is supported by DFG (FG 465).

\end{document}